# AN INVERSE MODELING METHOD TO ESTIMATE UNCERTAIN SPATIAL CONFIGURATIONS FROM 2D INFORMATION AND TIME-BASED VISUAL DISCRIMINATIONS

Pierre CUTELLIC[1]
[1]*ETH Zürich.*
[1]*cutellic@arch.ethz.ch, ORCID: 0000-0002-7224-9222*

**Abstract.** This paper focuses on a specific aspect of human visual discrimination from computationally generated solutions for CAAD ends. The bottleneck at work here concern informational ratios of discriminative rates over generative ones. The amount of information that can be brought to a particular sensory modality for human perception is subject to bandwidth and dimensional limitations. This problem is well known in Brain-Computer Interfaces, where the flow of relevant information must be maintained through such interaction for applicative ends and adoption of use in many fields of human activity. While architectural modeling conveys a high level of complexity in its processes, let alone in the presentation of its generated design solutions, promises in applicative potentials of such interfaces must be made aware of these fundamental issues and need developments of appropriate sophistication. This paper addresses this informational bottleneck by introducing a method to retrieve spatial information from the rapid serial visual presentation of generated pictures. This method will be explained and defined as inverse modeling, based on inverse graphics, and its relation to human visual processing.

**Keywords.** Neurodesign, Design Computing and Cognition, Brain-Computer Interfaces, Generative Design, Computer Vision.

## 1. Introduction

Human capacities in natural communication that find applicative echoes to remote intervention, rationalization, heuristics, and adaptivity in the face of complexity and uncertainty, have unparalleled performances compared to any state-of-the-art artificial model (Korteling et al., 2021). Stochastic problems, on the other hand, are legion in the industry, and the practical modeling of artificial intelligence has been proven to be of significant research interest across the spectrum of AEC fields (Mohammadpour and Asadi, 2019; Darko et al., 2020; Pan and Zhang, 2021). Virtuous coupling of both human and machine intelligence in guiding the diversity of sought performances in AEC technologies supports the right approach for more desirable outcomes, fluency, and literacy in both control and communication. However, numerous questions and bottlenecks exist at a general level of this hybrid model. They need to be answered in due time of research, both theoretically and practically. This paper focuses on a specific aspect of human visual discrimination from computationally generated solutions. The bottleneck at work here concern informational ratios of discriminative rates over generative ones. The amount of information that can be brought to a particular sensory modality for human perception is subject to bandwidth and dimensional limitations (Wickens, 1974; Venturino and Eggemeier, 1987; Klingberg, 2000; Marois and Ivanoff, 2005; Riener, 2017)—exceeding these rates cause well-known fatigue and disengagement. They can be measured in perceived mental efforts at various levels from physiological and behavioral signals as a cognitive load (Sweller and al., 1998; Xie and Salvendy, 2000; Paas and al., 2003). This might lead to counterproductive effects when the speed and quantity of solutions produced by a computer cannot be processed at a similar pace to their generation through human-computer interaction. This particular problem is well known in the field of Brain-Computer Interfaces (BCI), where the flow of relevant information must be maintained through such interaction for applicative ends and adoption of use in many areas of human activity (Lotte and Jeunet, 2018; Perkidis and Millan, 2020; Gramann and al., 2021). While architectural modeling conveys a high level of complexity in its processes, let alone in the presentation of its generated design solutions, promises in applicative potentials of such interfaces cannot escape from these fundamental issues and need developments of appropriate sophistication. Based on previous research (Cutellic, 2019, 2021) and the case study of an ongoing project tailoring BCI methods for architectural modeling at early design phases, this paper will address this informational bottleneck by introducing a technique to retrieve spatial information from the Rapid Serial Visual Presentation (RSVP) of generated pictures (Spence and Witkowski, 2013; Lees et al., 2018). The visual task at hand consists of focusing on sequences of generated pictures and discriminating visually those which seem relevant to the concept of a "ROOM" (or loosely describing an enclosing space). The visual discrimination is then further correlated with peculiar neural phenomena of study within the computational loop of a BCI (Figure 1). While the specifics of this experiment and the overall research project lie mainly outside of the scope of this paper, the intent of specific interest here is to identify presented pictures that may express stronger correlations with this concept and use them as sources of spatial information for subsequent designs.

# AN INVERSE MODELING METHOD TO ESTIMATE UNCERTAIN SPATIAL CONFIGURATIONS FROM 2D INFORMATION AND TIME-BASED VISUAL DISCRIMINATIONS

These pictures are necessarily framed by three constraints defined by minimal spatial cues: the monocular viewpoint to which they are presented, the aforementioned informational transfer rates, and maintaining a high degree of uncertainty to increase variance in the results of such interaction. The apparent trade-off that will be discussed throughout the presented method in this paper arises when one intends to model 3D information from insufficient and partial ones in 2D. The increased degree of uncertainty comes with increased intractability in modeling spatially complex 3D aggregates from 2D pictures containing low-level features. This method will be explained and defined as an inverse modeling method based on inverse graphics and their relation to human visual processing. Such a model of reference is usually called Vision-as-Inverse-Graphics (Kulkarni and Whitney, 2015) or scene Analysis-By-Synthesis (Grenander, 1976, 1978), where prior geometric information generally serves as ground truth. We will focus here on the primary input information only available by what is shown to the eye, and 3D correlates do not exist as priors.

## 2. Background

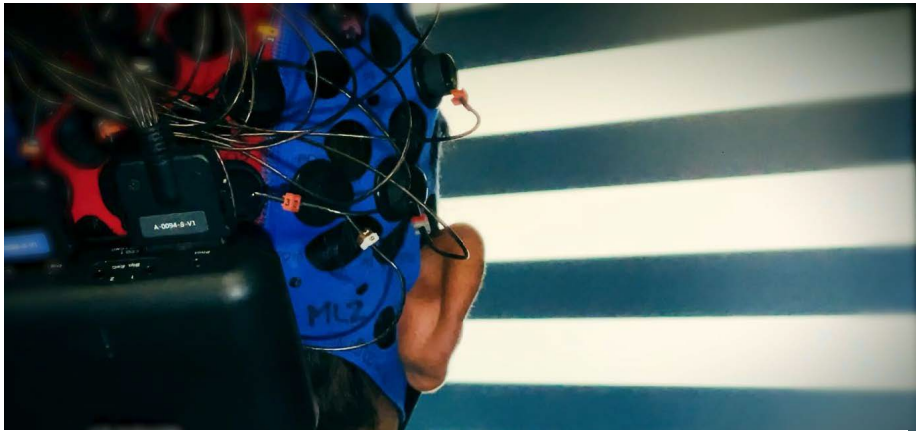

Figure 1. Photo of the conducted BCI experiment and its designed RSVP task for visual discrimination of spatial configurations.

The study of early visual processes seeks to model fast and efficient information recovery involved with cognitive functions of higher levels, such as object detection (Tomasi, 2006). One of the most basic and well-known feature extraction processes performed by the early visual system, and with computational equivalence, must deal with the segmentation of textures to extract shape information (Gibson, 1950; Marr et al., 1982). The so-called Shape From Texture problem in vision research focuses on modeling the recovery of 3D information of object surfaces from distorted texture segments which have been projected with specific angles onto a 2D Picture Plane, such as the retinal one (Gibson, 1950). This distortion, or Gradient, is generally measured by assuming the pattern distribution across the texture segment, whether a non-distorted reference is provided. Without reference, stochastic methods must be applied (Kanatani and Tsai-Chan, 1989; Clerc and Mallat, 2002). In the case of low-level and abstract geometry processing, such as for planar surfaces, distortions can be assumed to be linearly distributed and homogeneous. This assumption becomes even more practical when projections are considered within a Euclidean visual space because of the linear kind of its transformations (Erkelens, 2017), and deterministic methods can then be applied. From a perceptual point of view, perspective projection models are considered the most ideal, convenient, and realistic compared to transformations applied in physical and visual ones, which may depict both distance and size estimation in monocular vision with invariance (Erkelens, 2017). When a texture contains a dominant oriented structure, perspective convergence can be defined, and a mathematical relationship can be established between that 2D convergence in the texture and the parametric rotations of a planar surface in 3D by exploiting local symmetries (Saunders and Knill, 2001). Such convergence in texture gradients has been shown to represent further effective perceptual spatial cues for the extended generalization of ruled surfaces (Andersen, 1998; Gilliam, 1968; Todd et al., 2005). Given the abstract character of the visual objects to be detected in the designed experiment (a room), sets of abstract and basic geometry, such as planar surfaces, represent the most practical approach to describing 3D spatial segments. Using the perspective projection model, their projected distortion can be assumed uniform and reflected as homogeneous pattern distributions. Since only simple linear deformations are supposed to be present in the generated visual stimuli, deterministic numerical estimations of inverse monocular perspective projections of the deformed planar surfaces will be used.

## 3. Methods

Procedural methods that convey visuospatial information with adequate geometry must be developed to provide a serial and tractable way to model randomly generated pictures compound of such texture gradients. It is assumed that given a picture domain, providing a projection plane, and a singular viewpoint, given by the observer position, all kinds of randomly positioned and rotated planar surfaces defining a spatial enclosure within a field of view would produce straight intersections that delimit the viewed portions of these planes, and that such projection on the picture domain would map to the geometric properties of Voronoi space partition diagrams (Voronoi, 1909; Aurenhammer and Klein, 2000). Accordingly, a generative 2D domain is set to provide for the subsequent multiple texture segments that compose the picture (Figure 2). To deduce the segments from the diagram, all diagram cells must be closed, and their centroids must be randomly generated within an inset of the picture domain and an offset of the previously generated points for perceptually visible boundaries within the picture domain.

# AN INVERSE MODELING METHOD TO ESTIMATE UNCERTAIN SPATIAL CONFIGURATIONS FROM 2D INFORMATION AND TIME-BASED VISUAL DISCRIMINATIONS

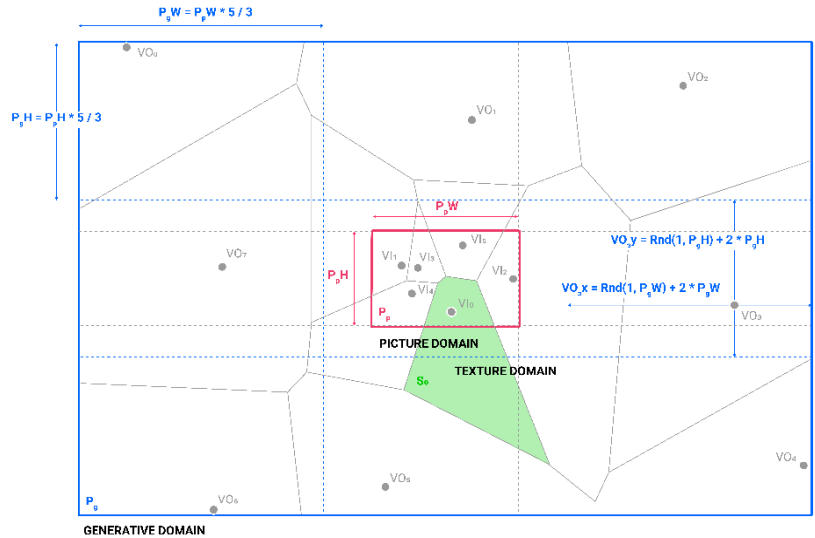

Figure 2. Relative generation of texture domains from a given 2D picture domain (Pp in red). A generative domain is set (Pg in blue) with a ratio to Pp width and height (PpW, PpH). Points are randomly placed within Pp (VIn) together with 8 points outside in their respective subdomains of Pg (VO0-7) to produce a closed Voronoi diagram containing all VIn. The resulting diagram provides for texture domains at each of its cells (S0 in green).

A most minimal texture gradient in the form of alternating black and white bands of equal dimensions (i.e., Square-Wave grating) is then taking advantage of the convex nature of the produced Voronoi cells and their bounding trapezoids (Figure 3). To generate the frequency bands, only one random integer parameter arbitrarily ranging from 1 to 10 is used with a similar distribution strategy to generating the coordinates of points on a plane. Another binary random integer is used to define whether the first band is to be black or white, and their orientation is predefined by using the boundary of each region. Both random cell distribution and texture frequencies are done to account for variance in the produced image. The longest edge of each area and its largest opposing edge (i.e., with at least one other edge separating them and the most parallel) are used to orient the texture bands. Most of the time, the two selected edges are not parallel, and each band may linearly vary its width. If a region contains only three edges, the opposing edge is generated by offsetting the largest edge to the opposite vertex and recreating an area with at least four vertices. The generated frequency bands remain bounded by the Voronoi cell and mapped onto the 2D picture with binary luminance values.

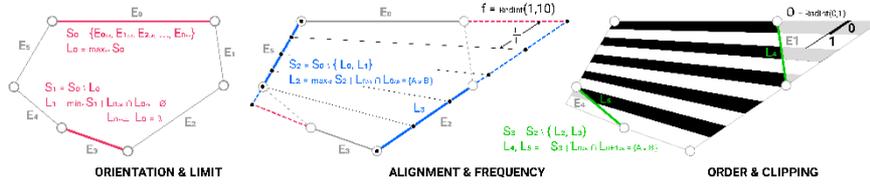

Figure 3. Selection of the longest edge (L0) and its opposing most-parallel edge (L1) to define orientation and limit of the texture. Middle: Selection of the longest connected edge (L2) and its opposed (L3) for each of the two previously selected to divide them by a generated frequency parameter (f) and define the alignment of the bands. Right, random order of the bands (O) by starting as 0-1 or 1-0. The remaining edges (L4, L5) form connected sets to clip the generated bands within the texture domain.

The procedural logic that defines the generation of texture segments on the picture plane is made to be reproducible and non-destructive. Given the same sets of geometrical elements, the same texture segments must be generated. Therefore, it must also account for rare cases of singularities where more than one element of the set may fit selection criteria and then apply further discrimination from their distinctive sign coordinates from the picture center. Since this procedure is aimed at generative modeling studies of increasing complexity, this principle must also be maintained across iterative sequences. A generator encodes the procedural generation of these pictures, and each generative information is stored in a dataset for retrieval and further modeling. Eventually, a token may serve as the seed for the generation of new texture segments with additional random points updating the initial Voronoi diagram along the iterative sequence. A simple domain union is made between the new and existing texture segments. That way, previously generated texture segments are preserved, and only their clipping boundaries may be updated regardless of whether the edges initially used for defining the texture have been altered (Figure 4).

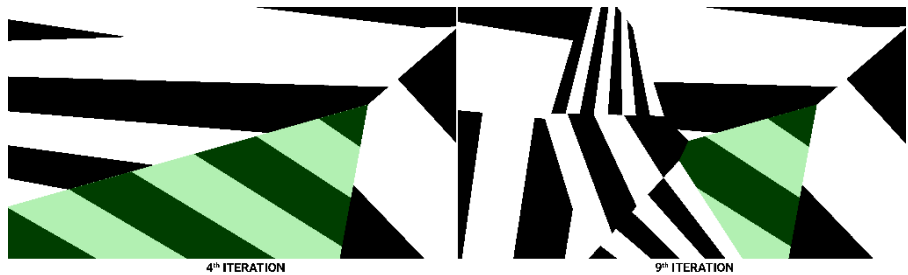

Figure 4. Two samples of generated images with texture segments through an iterative sequence. Left: the image generated after 4 iterations contains 4 texture segments. Right: the same image iterated over after 9 iterations. Highlighted in green is the same texture segment being updated over iterations.

# AN INVERSE MODELING METHOD TO ESTIMATE UNCERTAIN SPATIAL CONFIGURATIONS FROM 2D INFORMATION AND TIME-BASED VISUAL DISCRIMINATIONS

To recall the initial problem of inverse modeling from a monocular viewpoint (Palmer, 1999; Zygmunt, 2001): for a given surface projected on a 2D plane, there is an infinite number of possible configurations in the 3D field of view (Figure 5). The missing information allowing to position and orient a plane in a 3D domain is the distance from the viewpoint and Tilt and Slant rotation angles. Slant is the angle between the line of sight and the normal surface vector measured perpendicular to the picture plane, and tilt is the rotation angle of the normal surface vector around the line of sight (Stevens, 1983). Texture gradients provide for that missing information to perceive an uncertain but most probable surface orientation and position from a distance (Figure 6). As previously introduced, the geometric information provided by a texture segment on the picture plane, which also serves as perceptual information, is a function of the texture distribution. Segment-wise, local symmetry can be exploited to extract the slant and tilt angles. Two vanishing points may be retrieved from the longest edges of the trapezoid orienting the pattern generation, and their opposing ones generating its distribution (i.e., respectively L0, L1 and L2, L3 in Figure 3). Their bisector intersects at the center of transformations to be applied on the planar surface; the origin of the surface normal vector projected on the picture plane. Following (Ribeiro and Hancock, 1999), a linear approximation of the local distortion due to perspective can be derived. The projected normal vector components (p, q) can be retrieved by measuring the angle between the bisectors and the XY axes of the picture plane. From there, slant and tilt angles can be found. To serve as ground truth, a reference measure is used as the biggest possibly perceived pattern from a picture (half of the diagonal of the presentation screen). That reference is put in the Thales reciprocity formula with the measure of the distorted pattern along the bisector produced by the trapezoid edges generating the pattern distribution. The resulting distance is used for translating the rotated surface along the line of sight relatively to each other texture segments. The resulting planes must then be extended to fit with their visible boundaries punctuated by the extended line of sight passing through every vertex of the texture segment on the picture plane. The monocular perspective may only assume that a plane that is behind another may be occluded and then is extended until meeting the orthogonal projection of the boundary of the picture plane or simply intersecting with another plane. Similarly, a possibly occluding plane must stop at its sight boundaries for an occluded segment to be seen.

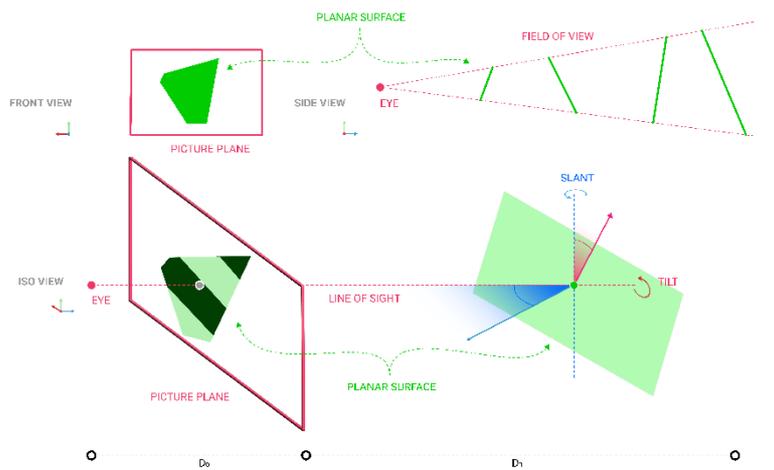

Figure 5. As seen from the eye (point in red), any projected surface (in green) on the 2D picture plane (as seen from the front view - top left) may actually be projected from an infinite number of possible configurations within the 3D field of view (as seen from the side view - top right). The three degrees of freedom which might be assessed from texture gradients on the picture plane to fix a plane in 3D are the slant (in blue) and tilt (in red) angles of the plane and the distance (D1 in black) from the picture plane to add with the eye distance from the picture plane (D0 in black), as seen from iso view - bottom.

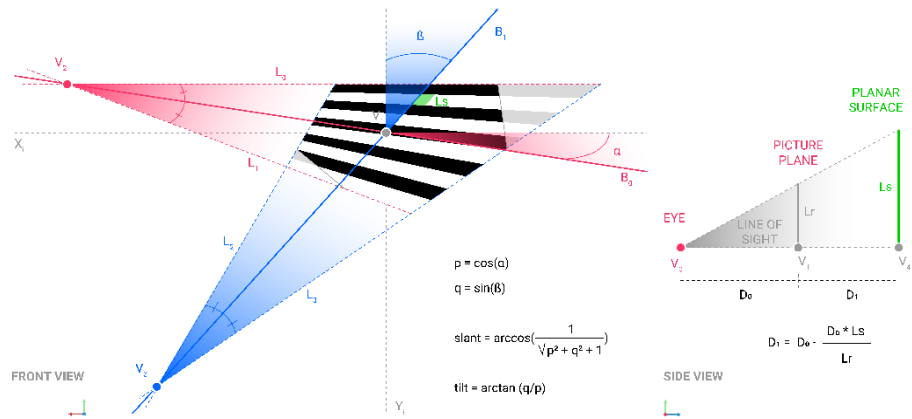

Figure 6. Left, front view, in the picture plane. The point of transformation V1 is retrieved by intersecting the bisectors B0 and B1. B0 and B1 pass respectively by the vanishing points V2 and V3 retrieved from the trapezoid edges generating the texture (L0, L1 and L2, L3). The linear equation that approximates the slant and tilt angles can be retrieved by measuring the angles alpha and beta between the bisectors and the axes of the picture plane. A distance measure of the texture Ls is taken along B1. - Right, side view, in the plane of the line of sight passing from the viewpoint origin V0 and V1. The distance D1 along this line to position the plane at V4 is deduced from the reference measure Lr by reciprocity.

## 4. Results

For reproducibility of the model under variance, random samples have been taken in the dataset of generated pictures (Figure 7). Since similar transformations must be found for identical texture segments regardless of their boundary conditions and appearance in the iterative process, the samples have been taken for different steps of the same sequence. Since the model must also be able to recover these transformations across the entire range of textures possibly generated under these design principles, samples have been taken for miscellaneous sequences. For every transformed sample, exactly similar transformations have been found for identical texture segments with different boundaries and were found to share the same plane in the inverse 3D space.

# AN INVERSE MODELING METHOD TO ESTIMATE UNCERTAIN SPATIAL CONFIGURATIONS FROM 2D INFORMATION AND TIME-BASED VISUAL DISCRIMINATIONS

The transformations were equally successfully applied to every other configuration presented in other generated sequences and matched perceptual evaluation of their orientation and relative distance as per the developed method. As expected, plane intersections vary greatly across generated samples. Since information regarding boundary conditions cannot be retrieved from the pictures, or if so, could be found in conflict with the present information regarding planar transformations, the former is considered to result from the latter. The retrieved boundary conditions become either the edge limiting a planar surface occluding geometries on the background and leaving a gap in-between, or planar intersections along a line or a point. From these resulting conditions, procedural geometric operations have been deduced and systematically applied to generate a CAD model and fabricate a physically tangible mock-up. Depending on the degree of control involved at this modeling stage, these features could be discussed further for designing pre-specific constraints or their parametrization for subsequent design explorations.

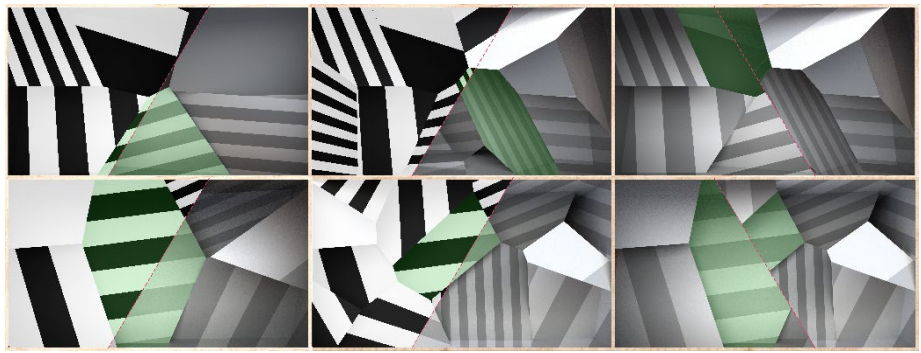

Figure 7. Two random samples of 2D generated pictures and their 3D planar geometries transformed through the inverse model (top and bottom). Columns left and middle show a comparison between 2D/3D for the corresponding generative steps 6 and 12. The third column (right) shows a comparison between 3D surfaces from both steps of the same generative sequence. Highlighted in green is a corresponding texture patch for a given sequence and across steps. For reasonable productions, surfaces have been bounded to a standard enclosing box and onto which trimmed surface regions have projected for visual continuity along the monocular perspective.

## 5. Conclusion and Outlook

This paper presents a method to design an inverse geometric model of 3D planar geometries from 2D perceptual information under a certain degree of uncertainty and for the rapid presentation of early design scenarios. Its low level of 2D information and presentation pace aims to develop applicative methods in BCI for design and architectural modeling that support the increase of fluency between humans and computers. Followingly, minimal and basic 2D texture information, which does not necessitate large amounts of time to generate procedurally, has been used to model the retrieval of 3D information. The core of such an approach lies in two basic assumptions of productivity driven by cumulative empirical findings from cognitive science literature: the generative productivity of computers to output solutions from large dimensional spaces and the discriminative productivity of humans to output responses of higher dimensions from reduced ones. Intractable problems due to a lack of sufficient and timely information may be approached heuristically and iteratively. Several assumptions have been made during the development of this method and kept here for discussion because their relevance mainly concerns the degree of generalization for further research. The basic assumption which allows for correlating a geometric inverse model with perceptual cues examines the hypothesis that regardless of the complexity of visual inputs (i.e., natural images), the principle of inverse geometric modeling operates naturally at the level of human perceptual segmentation. In addition, monocular models are known to leave out a significant amount of uncertainty. For example, a perceived slant angle may vary from two different instances of the same texture segment (Blake, Bülthoff, & Sheinberg, 1993; Cutting & Millard, 1984; Knill, 1998a). The estimation error in perceived angles may also vary by angle thresholds and pattern differences. Additionally, the statistical nature of such information is related to the presentation context (e.g., the overall composition of all texture segments, the modality from person to person, …). Because RSVP integrates by design the repetitive presentation of similar instances on an individual basis, and parameters of uncertainty in the perception of texture gradient are continuously accumulated in the literature, a certain degree of confidence could be further developed for these 3D transformations. Regarding the remaining uncertainty parameters regarding relative distance estimations and surface occlusions, further investigation in providing additional viewpoints could give the method more accuracy. Optimal cue combination methods can support this development, and the deployment of such a process could be thought either stationary (presentation of multiple viewpoints of the same scene on a static screen) or mobile (in the context of augmented reality devices). Finally, generated texture gradients should account for visual discomfort under prolonged presentation modalities such as RSVP. Artificially generated distributions could mitigate this effect by eventually accommodating noise found in natural images. The generation of such noise-based gradients would involve applying generalized methods, including stochastically distributed patterns and non-planar surface geometries.

## Acknowledgments

This research is part of the Neuramod project (105213_192500), fully funded by the Swiss National Science Foundation SNSF Project Funding Div. 1-2 for Humanities, Social sciences, Mathematics, Natural, and Engineering sciences.

# AN INVERSE MODELING METHOD TO ESTIMATE UNCERTAIN SPATIAL CONFIGURATIONS FROM 2D INFORMATION AND TIME-BASED VISUAL DISCRIMINATIONS